\documentclass{article}
\usepackage{spconf,amsmath,graphicx}
\usepackage{booktabs}
\usepackage{multirow}
\usepackage{cite}
\usepackage[colorlinks,linkcolor=red,anchorcolor=blue,citecolor=green,CJKbookmarks=True]{hyperref}       % hyperlinks

\title{MULTI-TASK AUDIO SOURCE SEPARATION}
\name{Lu Zhang$^{1,2}$, Chenxing Li$^1$, Feng Deng$^1$, Xiaorui Wang$^1$}
\address{$^1$Kuai Shou Technology Co., Beijing, China\\
         $^2$Harbin Institute of Technology, Shenzhen, China}
\begin{document}
%\ninept
%
\maketitle
\begin{abstract}
The audio source separation tasks, such as speech enhancement, speech separation, and music source separation, have achieved impressive performance in recent studies. The powerful modeling capabilities of deep neural networks give us hope for more challenging tasks. This paper launches a new multi-task audio source separation (MTASS) challenge to separate the speech, music, and noise signals from the monaural mixture. First, we introduce the details of this task and generate a dataset of mixtures containing speech, music, and background noises. Then, we propose an MTASS model in the complex domain to fully utilize the differences in spectral characteristics of the three audio signals. In detail, the proposed model follows a two-stage pipeline, which separates the three types of audio signals and then performs signal compensation separately. After comparing different training targets, the complex ratio mask is selected as a more suitable target for the MTASS. The experimental results also indicate that the residual signal compensation module helps to recover the signals further. The proposed model shows significant advantages in separation performance over several well-known separation models.

\end{abstract}
%
% Section 1: Introduction
\section{Introduction}
\label{sec:intro}
With the development of online live broadcasts and short videos, the demand and difficulty of processing front-end audio signals are also increasing in the era of mobile Internet. The audio signals recorded in these scenes usually contain various types of sources, such as speech, music, and background noises. These sources may overlap and obscure each other. In such scenarios, the signals of speech and music are equally important. In the background noises, there are also some acoustic events that we are interested in. In addition, estimating background noise is very useful for evaluating the audio quality or enhancing speech signals. This brings a new research challenge called multi-task audio source separation (MTASS), that is, how to separate speech, music, and noise from the mixture.

Recent years we have witnessed the remarkable progress of deep neural networks (DNNs) in the study of audio source separation. Previous works mainly focus on the single-task audio source separation, such as speech enhancement \cite{R2,R3,R1} and speech separation \cite{R4, R5, R7}. The purpose of speech enhancement is to recover high-quality speech signals from the mixed signals of noise and speech, while the speech separation task aims to separate the speech of different speakers. Due to the different characteristics of the separated target signal, the two tasks also differ in the choice of network structure and signal analysis domain. Recent studies \cite{R8,R9,R10} have shown that the complex domain of short-time Fourier transform (STFT) is more suitable for enhancing speech from background noises, and the learned time-domain analysis and synthesis transforms works better for speech separation \cite{R7,R11}. A variety of network structures has been successfully applied to these two separation tasks, including long short-term memory (LSTM) recurrent network \cite{R3,R5,R12,R13}, convolutional recurrent network (CRN) \cite{R6,R8,R9}, temporal convolutional network (TCN) \cite{R7,R14}, and transformer-based network \cite{R15, R16, R17}. This also shows that capturing the long-term dependence of the audio signal is crucial for better separation performance. 

As another audio source separation tasks, music source separation \cite{R18,R19,R20,R21,R23,R22} has long been an important topic in music information retrieval. Its purpose is to separate different sound source components from a music recording, such as vocal, guitar, drum, and other musical instruments. A classic open-source system \cite{R20}, called Open-Unmix, applies a bidirectional LSTM network for separating different music sources one by one in the frequency domain. A U-Net-LSTM-based model, Demucs \cite{R21}, directly operates on the time-domain waveform and obtains state-of-the-art performance. \cite{R22} considers both time-domain and frequency-domain information and defines a combined loss to constrain the learning of the model to prevent source leakage. Very similar to the music source separation, the singing voice separation \cite{R24,R25,R26} task only needs to separate the singing voice and the musical accompaniment, which has also received much attention. Learning the complex ratio masking (cRM) in a self-attention, Dense-UNet \cite{R26} addresses the importance of phase and obtains state-of-the-art results.

This paper divides the common audio sources into speech, music, and background noise to achieve a MTASS. Different from the universal source separation task in \cite{R27,R28,R29}, which separates all arbitrary sound sources in a piece of mixed audio, our task aims to separate the three fixed types of sound sources into three tracks. More specifically, the output of the speech track is a normal speaking voice, and the music track signal we define is a broad category, which may be full songs, vocals, and different accompaniments. Except for music and speech, any other possible background sounds, such as closing doors, animal calls, and some annoying noises, are classified as noise track signals. 

This paper investigates MTASS in-depth and provides three main contributions: (1) We first introduce the MTASS task and propose an open-source dataset in which mixtures containing three types of audio signals. (2) Learning the success of noise estimation in the frequency domain \cite{R30} and the frequency sparsity of the music signal itself \cite{R19}, we carefully select the training targets and perform a detailed comparison in MTASS. And then, we conduct the MTASS in the complex domain. (3) We propose a two-stage-based model for MTASS, named Complex-MTASSNet. It consists of a separation module and a residual signal compensation module, which first performs the separation and then refines the signals of each track. Experiments also demonstrate the advantages of our model compared with several well-known time-domain-based and frequency-domain-based models.

% Section 2: MTASS Dataset
\section{MTASS Dataset}
\label{sec:format}

In this section, we describe the construction of the MTASS dataset. First, we prepare three types of datasets for the generation of the mixed dataset. Table~\ref{tab:datag} lists the source datasets we selected for speech, music, and background noises. Two Mandarin datasets, Aishell-1 \cite{R31} and Didi Speech \cite{R32}, are used to build the speech source dataset. Aishell-1 has a total of 178 hours of speech data spoken by 400 people, including 340 people in the training set, 20 people in the test set, and 40 people in the validation set. The DiDi Speech dataset contains about 60 hours of speech data, spoken by 500 people. We randomly select 400 people as the training set and divide the data of the remaining 100 people into a test set and a validation set of 50 people each. The demixing secrets dataset (DSD100) of the Signal Separation Evaluation Campaign (SISEC) \cite{R33} is used as the music source dataset. We preprocess this music dataset\footnote{https://github.com/faroit/dsdtools} and obtain several isolated files of the full song, bass, drum, vocal, and other accompaniments. We randomly select 70 songs as the training set, 15 songs as a validation set, and the last 15 songs as a test set. For the noise source dataset, we directly use 181 hours of noise data from the Deep Noise Suppression (DNS) Challenge \cite{R34}.  We also divide the whole noise dataset into three parts: training set, validation set, and test set, of which 163 hours are training set, and both validation set and test set are 9 hours.

\begin{table}[t]\large
  \caption{The source datasets used for the generation of MTASS dataset.}
  \label{tab:datag}
  \centering
  \vspace{0.1cm}
  \resizebox{1\columnwidth}{!}{
    \begin{tabular}{ccccc}
    \toprule
     & Dataset & Data format & Sampling rate & Data duration   \\
    \midrule
    \multicolumn{1}{c}{\multirow{2}{*}{\text{Speech}}}
    & AiShell-1 &.wav &16Khz &178 hours    \\
    & Didi Speech &.wav &48Khz &60 hours    \\
    Music & DSD100 &.wav &44.1Khz &38 hours  \\
    Noise & DNS  &.wav &16Khz &181 hours \\
    \bottomrule
    \end{tabular} 
  }
\end{table}

Before creating the mixed audio, we first split the original speech, music, and noise data into 10-second segments. The speech and music data need to be resampled into 16Khz audio files and then segmented. After segmentation, a total of 86,101 speech audio clips are obtained, including training data (71,720), validation data (8614), and test data (5767). 12,500 music audio clips are also acquired, including training data (8781), validation data (1696), and test data (2023). The obtained noise clips are 36,003 in total, corresponding to the training data (29,337), the validation data (3333), and the test data (3333). Then, we randomly read data from the segmented speech, music, and noise files and mix them to obtain a dataset for MTASS. For each speech clip, music and noise clips are added with a random signal-to-noise (SNR) ratio of -5 to 5dB. In this study, we generate a total of 22,000 sets of audio data (mixed audio and ideal audio for each track), including 20,000 training data (55.6 hours), 1000 validation data (2.8 hours), and 1000 test data (2.8 hours). The details of the generated MATSS dataset are listed in Table~\ref{tab:MTASSdata}. The instructions and generation scripts for the MTASS dataset are also published\footnote{https://github.com/Windstudent/Complex-MTASSNet/}. 
% at \url{https://github.com/Windstudent/Complex-MTASSNet/}.

\begin{table}[t]\large
  \caption{The details of MTASS dataset.}
  \label{tab:MTASSdata}
  \centering
  \vspace{0.1cm}
  \resizebox{1\columnwidth}{!}{
    \begin{tabular}{ccccc}
    \toprule
     & Dataset & Data format & Sampling rate & Data duration   \\
    \midrule
    \multicolumn{1}{c}{\multirow{3}{*}{\text{MTASS}}}
    & train &.wav &16Khz &55.6 hours \\
    & validation &.wav &16Khz &2.8 hours \\
    & test &.wav &16Khz &2.8 hours \\
    
    \bottomrule
    \end{tabular} 
  }
\end{table}

% Section 3: Proposed Model 
\section{Proposed Model}
\label{sec:pagestyle}

\begin{figure*}[t]
\centering
	\includegraphics[scale=0.58]{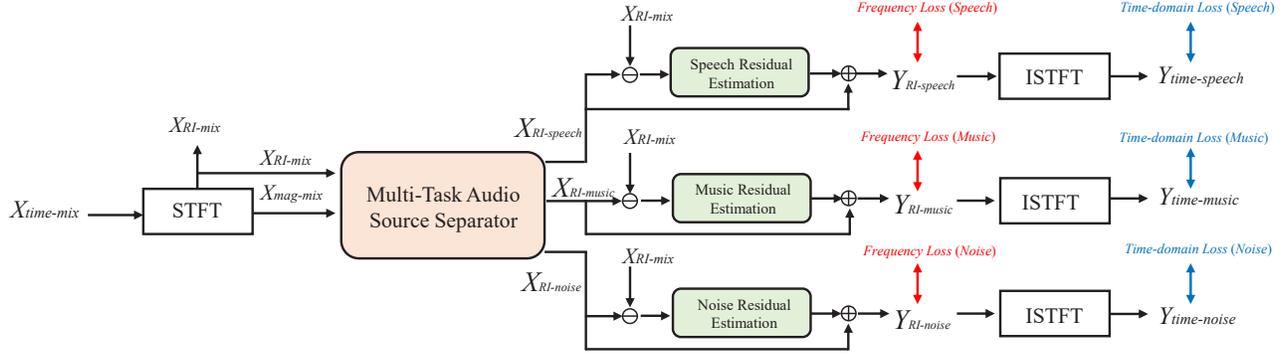}
\vspace{-0.1cm}
\caption{The detailed framework of Complex-MTASSNet.}
\label{fig:model}
\end{figure*}

To tackle this challenging multi-task separation problem, we also proposed a two-stage DNN model to separate the speech, music, and noise track signals, as shown in Fig.~\ref{fig:model}. Based on the analysis and synthesis transform of STFT, Complex-MTASSNet separates the signal of each audio track in the complex domain and further compensates the leaked residual signal for each track. The details are as follows.

\begin{figure*}[t]
\centering
	\includegraphics[scale=0.6]{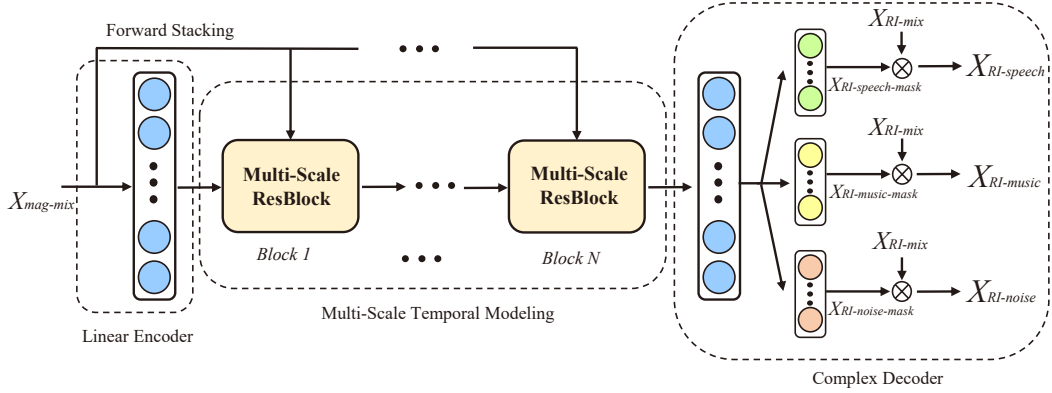}
\vspace{-0.1cm}
\caption{The block diagram of multi-task audio source separator.}
\label{fig:Complexmstcn}
\end{figure*}

\subsection{Multi-Task audio source separator}
\label{ssec:separator}

As shown in Fig.~\ref{fig:Complexmstcn}, the separator adopts a multi-scale TCN structure \cite{R14} and adds a complex decoder to separate different sources. It takes the magnitude spectrum of the mixture as input and produces three masks to perform complex-domain filtering. In the encoder part, the input feature is first transformed to a higher-dimensional representation through a fully connected linear layer. Then, stacking many multi-scale ResBlocks can model the temporal dependency of audio signals at a more granular level. The complex ratio mask of each track is estimated separately in the output part of the proposed separator. Considering that complex-domain signals have positive and negative properties, we use linear layers to learn complex ratio masks (cRMs) and define the mean square error (MSE) loss on the complex spectrum so that the model can implicitly learn a mask that is more suitable for separation:

\begin{equation}
\text { MSE }=\sum_{i=1,2,3}\left[X_{mask,i} * X_{RI, mix}-S_{RI,i}\right]^{2}
\end{equation}
where \(X_{mask,i}\) is an estimate of cRM, with index i=1,2,3, which respectively represent the speech, music, and noise tracks. \(X_{RI, mix}\) is the complex spectrum (real and imaginary spectrum) of mixture signals, and \(S_{RI,i}\) represents the ideal RI spectrum for each track.

In our model, we use STFT to analyze the time-frequency characteristic of mixture audio, where the window size and hop size are set to 512 and 256, respectively, and a total of 257 dimension magnitude spectrum is obtained. The dimension of the linear encoder is set to 1024, and the number of channels of three convolutional layers in the multi-scale ResBlock is 257-514-1024. Original magnitude spectrum is stacked forward into each ResBlock and concatenated with the extracted features to perform multi-scale temporal analysis of 8 subbands. In our separator model, we stack \emph{N} multi-scale ResBlocks, and the dilation rate cycles in increments of 1, 3, 5, 7, and 11. In the complex decoder, the dimension of the linear transformation layer is also set to 1024, and the output mask for each of the three tracks is 514 (257 for the real part and 257 for the imaginary part).

\subsection{Residual signal estimation module}
\label{ssec:residual estimator}

Although the separator can separate the signals of each audio track, each track may have residual signals leaked from other tracks. Therefore, we propose a two-stage processing structure that separates first and then compensates. The designed residual signal estimation module can estimate the residual signals of the target audio track from the non-target audio tracks and then compensate the target audio track. The process of residual estimation module can be given as:
\begin{equation}
X_{res, i}=F\left(X_{RI, mix}-X_{RI, i}\right)
\end{equation}
where \(X_{res, i}\) represents the estimated residual signal of the \emph{i}-th track, $\emph{F}(\cdot)$ represents the processing and transformation of residual signal estimation module. The final output of each track is obtained by adding the output of the separator and the estimated residual signal:
\begin{equation}
Y_{RI, i}=X_{RI, i}+X_{res,i}
\end{equation}

\begin{figure}[t]
\centering
	\includegraphics[scale=0.64]{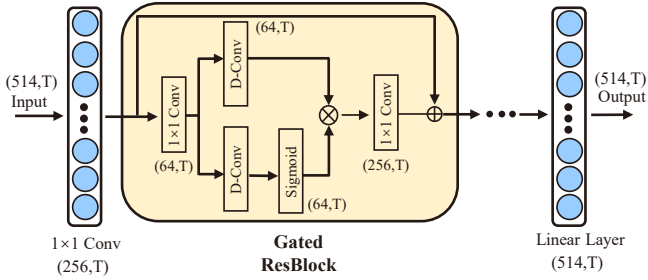}
% \vspace{-0.2cm}
\caption{The block diagram of Gated TCN model for the residual signal estimation.}
\label{fig:gatedTCN}
\end{figure}

For the design of the residual estimation model, we adopt a gated TCN structure, as shown in Fig.~\ref{fig:gatedTCN}. It should be noted that all the norm layers and activation layers are omitted for convenience, and they are all denoted as `D-Conv' in the figure. The regularization methods used here are batch normalization \cite{R35} and dropout \cite{R36}, which are added after the convolution operation. In this model, given input with size (514, T) where 514 and T denote the dimension of separated RI spectrum and the timesteps, respectively, we first compressed it to a 256-dimension feature using a 1×1 convolutional layer. The gated ResBlock is composed of an input 1×1 convolutional layer, two parallel dilated convolutional layers, and an output 1×1 convolutional layer. The input convolution compresses the channel number into 64 and is followed by dilated convolutions with a gating mechanism to allow interaction between branches. To meet the needs of residual connection, the output convolutional layer expands the features to 256 dimensions. In this model, \emph{M} gated ResBlocks with dilation factors 1, 2, 4, ..., $2^{M-1}$ are repeated \emph{R} times to model the long-term context information of target audio signals. The output of the model is a 514-dimensional linear layer to estimate the complex residual signals, \(X_{res, i}\).

\subsection{Loss function}
\label{ssec:loss function}

In our two-stage model, we adopt MSE as the loss function on the complex spectrum and combine the time-domain SNR loss to achieve multi-domain optimization. The multi-domain loss on the three-track signals is defined as follows:
\begin{equation}
\begin{aligned}
L(S, Y)=\sum_{i=1,2,3} &\left[\operatorname{MSE}\left(Y_{R I, i}, S_{R I, i}\right)\right.\\
&\left.+\alpha \cdot \operatorname{SNR}\left(y_{t i m e, i}, s_{\text {time }, i}\right)\right]
\end{aligned}
\end{equation}
\begin{equation}
\mathrm{SNR}=-10 \log _{10}\left(\frac{\sum_{t=1}^{T}\left\|y_{\text {time }, i}\right\|^{2}}{\sum_{t=1}^{T}\left\|y_{\text {time }, i}-s_{\text {time }, i}\right\|^{2}}\right)
\end{equation}
where \(y_{\text {time }, i}\) and \(s_{\text {time }, i}\) respectively represents the separated and ideal time-domain signals. \emph{T} represents the total frames in an utterance. The scaling parameter $\alpha$ is set to 0.01 to ensure the two parts of the loss of the same order of magnitude. Unlike other two-stage models, which require adding loss functions to both separate stages,  we optimize both components (separator and residual estimation) at the same time using only one loss function. This is because the final estimated output  \(Y_{RI, i}\) is obtained by directly adding the outputs of the first and the second stages so that the two components are more like a whole to optimize a common loss.

\begin{figure*}[t]

\centering
	\includegraphics[scale=0.65]{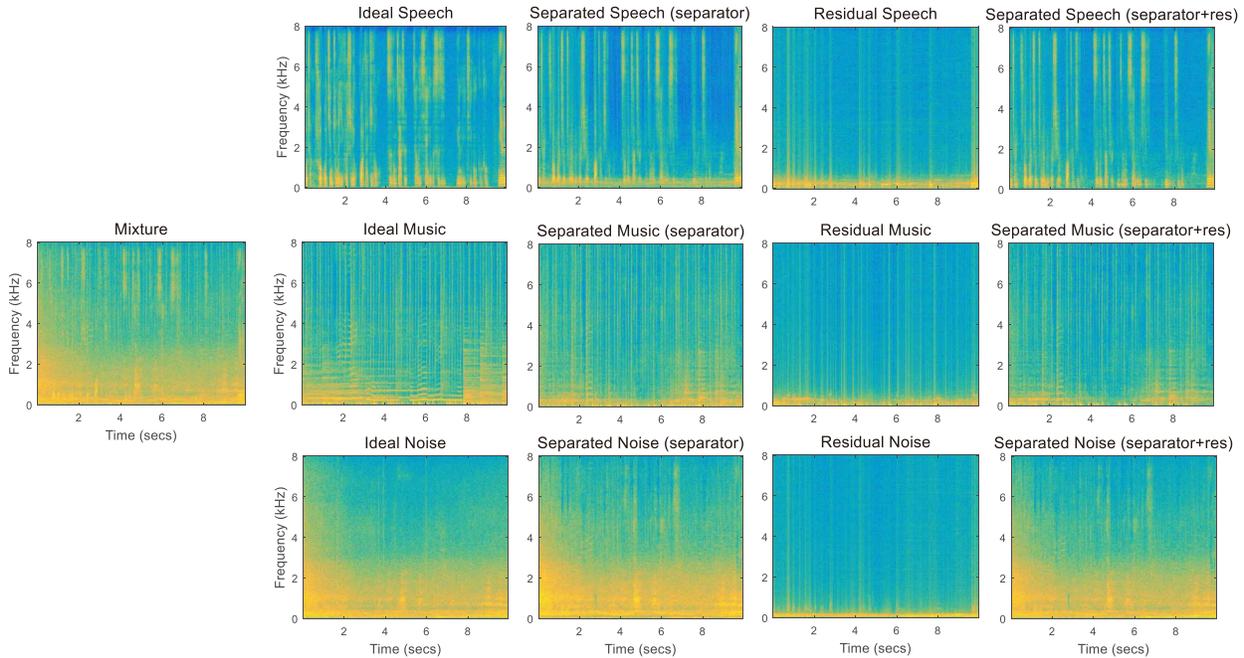}
\vspace{-0.1cm}
\caption{Spectrogram of the separated signals for different model stages.}
\label{fig:Spectrogram}
\end{figure*}

% Section 4: Experiments 
\section{Experiments And Results}
\label{sec:majhead}

\subsection{Experimental setup}
\label{ssec:subhead}

In our experiments, the proposed Complex-MTASSNet is performed using PyTorch and trained with the Adam optimizer \cite{R37}. The initial learning rate is set to $1 e^{-3}$ and halved if the loss of validation set is not improved. In our model, 8 gated ResBlocks are repeated 5 times for residual signal estimation. The kernel size of all dilated convolutions is set to 3. Separation performance is measured by the signal-to-distortion ratio (SDR) \cite{R38}, which is defined as:
\begin{equation}
\text { SDR }:=10 \log _{10} \frac{\left\|s_{\text {target }}\right\|^{2}}{\left\|e_{\text {interf }}+e_{\text {mise }}+e_{\text {artif }}\right\|^{2}}
\end{equation}
where $e_{\text{interf}}$, $e_{\text{noise}}$, and $e_{\text{artif}}$ are, respectively, the interferences, noise, and artifacts error terms.

\subsection{Comparisons of different training targets}
\label{ssec:ex_sess_1}

In addition to the model structure, training targets also play an important role in the supervised audio source separation. Based on the previous researches on training targets in speech enhancement and separation tasks, we first investigate the training targets in MTASS. The evaluation results for different training targets of the proposed separator are reported in Table~\ref{tab:difftar}. 

\begin{table}[th]
  \caption{SDR improvement (SDRi) for different training targets of the proposed separator obtained on the speech, music, and noise signal tracks.}
  \label{tab:difftar}
  \centering
  \resizebox{0.95\columnwidth}{!}{
  \begin{tabular}{ccccc}
    \toprule
    \multirow{2}{*}{Methods} & \multicolumn{4}{c}{\textbf{SDRi (dB)}} \\
                             &Speech           &Music              &Noise           &Ave\\
    \midrule
    Time (5 MS-ResBlocks)        &10.98            &7.76               &6.91            &8.55\\
    Mag (5 MS-ResBlocks)         &10.44            &7.86               &7.36            &8.55\\
    RI (5 MS-ResBlocks)          &10.42            &7.37               &5.96            &7.92\\
    cRM (5 MS-ResBlocks)         &10.56            &7.87               &7.54            &8.66\\
    \midrule
    cRM (10 MS-ResBlocks)        &10.84            &8.23               &7.89            &8.99\\
    cRM (15 MS-ResBlocks)  &\textbf{11.10}  &\textbf{8.32}      &\textbf{8.10}  &\textbf{9.17}\\
    \bottomrule
  \end{tabular}
  }
\end{table}

As shown in Table~\ref{tab:difftar}, the effectiveness of the time-domain-based and frequency-domain-based training targets are compared. The time-domain-based training target (denoted as `Time') is similar to the targets of Conv-TasNet \cite{R7}, which uses the 1-D convolutions to replace the STFT as the analysis tool and directly output the separated waveform. The other three targets, magnitude ratio mask (`Mag'), complex spectrum (`RI'), and cRM, are all frequency-domain-based representations obtained using STFT/ISTFT as the analysis-synthesis tool. From Table~\ref{tab:difftar}, we can observe that the model with the cRM-based target performs balance on the three audio tracks. Moreover, increasing the number of multi-scale ResBlocks is beneficial to improve the performance of the proposed separator.

\subsection{The effectiveness of two-stage model}
\label{ssec:ex_sess_2}

In this subsection, we evaluate the effectiveness of the proposed two-stage processing pipeline. Fig.~\ref{fig:Spectrogram} shows the output spectrogram of separator, residual estimation module, and the compensated signals in the Complex-MTASSNet. It can be found that the `separator+res' has less distortion than the output of `separation', which means that the estimated signal does help to compensate for some spectral details. Taking the speech track as an example, we can find that the residual interference of the separated speech after compensation is less than the output of the separator. It means that the residual compensation module also plays a role in suppressing the leakage of other audio tracks.

\begin{table}[th]
  \caption{SDRi for the proposed model with Time-domain loss (T-Loss) and Frequency loss (F-Loss) obtained on the speech, music, and noise signal tracks.}
  \label{tab:sep+res}
  \centering
  \resizebox{1.0\columnwidth}{!}{
  \begin{tabular}{ccccc}
    \toprule
    \multirow{2}{*}{Methods} & \multicolumn{4}{c}{\textbf{SDRi (dB)}} \\
                             &Speech           &Music              &Noise           &Ave\\
    \midrule
    1 stage+F-Loss &11.10         &8.32            &8.10            &9.17\\
    \midrule
    2 stage+F-Loss (separator-out)  &9.85          &6.63              &7.73            &8.07\\
    2 stage+F-Loss (separator+res-out) &12.13         &9.39     &\textbf{8.47}            &10.00\\
    2 stage+TF-Loss (separator-out) &10.22         &6.17              &7.24            &7.88\\
    2 stage+TF-Loss (separator+res-out) &\textbf{12.57} &\textbf{9.86}    &8.42 &\textbf{10.28}\\
    \bottomrule
  \end{tabular}
  }
\end{table}

Table~\ref{tab:sep+res} presents the experimental results of the Complex-MTASSNet using the proposed multi-domain loss function. It can be observed that the model combining time-domain and frequency loss (TF-Loss) performs better than the model using only frequency loss (F-Loss). This performance advantage is obvious in the speech and music tracks. No matter which loss function is used, the residual signal estimation module can improve the speech distortion of the output of each track of the separator. We also find an interesting phenomenon that the SDRi of the separator obtained by joint training is inferior to that of the separator trained alone (1 stage+F-Loss). However, the performance of the two-stage model after compensation will be significantly improved.

\subsection{Comparisons with other models}
\label{ssec:ex_sess_3}

In order to demonstrate the superiority of the proposed model in this multi-task separation, it is necessary to compare Complex-MTASSNet with other related methods. In this section, we have evaluated the separation performance and computational complexity of the Complex-MTASSNet and compared it with several well-known baselines in speech enhancement, speech separation, and music source separation, which are GCRN\footnote{https://github.com/JupiterEthan/GCRN-complex} \cite{R8}, Conv-TasNet\footnote{https://github.com/naplab/Conv-TasNet} \cite{R7}, Demucs\footnote{https://github.com/facebookresearch/demucs} \cite{R21}, and D3Net\footnote{https://github.com/sony/ai-research-code/tree/master/d3net} \cite{R23}. 

\vspace{-0.2cm}
\subsubsection{Separation performance}
\label{ssec:ex_subsess_1}

Table~\ref{tab:compmodel} reports the SDRi of different models on three audio tracks. For baselines, we modify the number of outputs into three, corresponding to the three tracks of speech, music, and noise. `GCRN-RI' and `GCRN-cRM' respectively represent the GCRN model whose learning targets are complex spectrum and cRM.

\begin{table}[th]
  \caption{SDRi for the comparison models obtained on the speech, music, and noise signal tracks.}
  \label{tab:compmodel}
  \centering
  \resizebox{0.90\columnwidth}{!}{
  \begin{tabular}{ccccc}
    \toprule
    \multirow{2}{*}{Methods} & \multicolumn{4}{c}{\textbf{SDRi (dB)}} \\
                             &Speech           &Music              &Noise           &Ave\\
    \midrule
    GCRN-RI \cite{R8}        &9.11          &5.76              &5.51            &6.79\\
    GCRN-cRM \cite{R8}       &8.73          &6.25              &6.50            &7.16\\
    Demucs \cite{R21}        &9.93          &6.38              &6.29            &7.53\\
    D3Net \cite{R23}         &10.55         &7.64              &7.79            &8.66\\
    Conv-TasNet \cite{R7}    &11.80         &8.35              &8.07            &9.41\\
    Complex-MTASSNet &\textbf{12.57}   &\textbf{9.86}        &\textbf{8.42}  &\textbf{10.28}\\
    \bottomrule
  \end{tabular}
  }
\end{table}

From the Table~\ref{tab:compmodel}, as for the frequency-domain-based methods, GCRN-cRM performs better than GCRN-RI, which also proves the superiority of cRM in multi-task separation. GCRN-cRM performs inferior than the 1-stage-based Complex-MTASSNet (models listed in Table~\ref{tab:difftar}) and the proposed Complex-MTASSNet, which shows the advanced nature of the structures in the MTASS. Conv-TasNet performs best among baselines, which mainly attributes to the structures and the time-domain-based end-to-end training. The proposed two-stage strategy and the multi-domain loss effectively help Complex-MTASSNet to obtain performance improvement. In general, Complex-MTASSNet achieves the best SDRi on all three tracks, which are 12.57dB, 9.86dB, and 8.42dB on speech, music, and noise track, respectively.

\vspace{-0.2cm}
\subsubsection{Model size and complexity}
\label{ssec:ex_subsess_2}

In addition to separation performance, we also pay attention to the size and computational complexity of the model, which are also key points that need to be considered when deploying the model in the application scenarios. Model size (parameters), multiply-accumulate operations per second (MAC/S), and the processing time consumption per second on CPU (Intel Xeon 5120) and GPU (Nvidia 2080Ti) are all concerned, and the detailed results are listed in Table~\ref{tab:complexity}.

We can find that the parameters of Complex-MTASSNet is 28.18 M, which is only smaller than Demucs, but the proposed model requires the least amount of computation. This is because Complex-MTASSNet mainly uses 1-D convolutions, which avoids the computational burden caused by using a large number of shared convolution operations. From the real-time tests, the proposed model shows good real-time performance on both CPU and GPU platforms. Experimental results show that the Complex-MTASSNet performs relatively balanced on complexity tests while ensuring the best separation performance.

\begin{table}[t]
  \caption{Model parameters, MAC/S and the processing time consumption per second (real-time factor, RTF) of the comparison models.}
  \label{tab:complexity}
  \centering
  \resizebox{1.0\columnwidth}{!}{
  \begin{tabular}{ccccc}
    \toprule
    Models         &Parameters    &MAC/S    &RTF (GPU)   &RTF (CPU)\\
    \midrule
    GCRN  \cite{R8}          &9.88 M       &2.5 G     &0.031    &0.230\\
    Demucs \cite{R21}        &243.32 M     &5.6 G     &0.006    &0.288 \\
    D3Net \cite{R23}         &7.93 M       &3.5 G    &0.002    &0.139\\
    Conv-TasNet \cite{R7}    &5.14 M       &5.2 G     &0.017    &0.836\\
    Complex-MTASSNet         &28.18 M    &1.8 G     &0.019    &0.391\\

    \bottomrule
  \end{tabular}
  }
\end{table}

% Section 5: Conclusion 
\section{Conclusion}
\label{sec:conclu}

This paper introduces a new MTASS problem and constructed a dataset of mixtures containing speech, music, and background noises. In response to this challenging problem, we propose a processing idea of separating first and then compensating and develop a two-stage model based on TCN. By exploring different training targets, cRM is more suitable for this task. Compared with the direct separation method, the two-stage processing strategy shows significant advantages in separation performance. Our experiments indicate that the proposed Complex-MTASSNet is not only superior in separation performance to the well-known models of other separation tasks, but also has lower computational complexity. The codes, the pre-trained models, and more analysis will be released\footnote{https://github.com/Windstudent/Complex-MTASSNet/}. We hope the proposed MTASS task and Complex-MTASSNet can give researchers more inspiration, and the proposed dataset can lower the barrier to this new research.

\bibliographystyle{IEEEbib}
\bibliography{refs}

\end{document}